\documentclass[aps,prl,twocolumn,superscriptaddress,byrevtex,showpacs,preprintnumbers,amsmath,amssymb,floatfixm]{revtex4}
\usepackage{wasysym}
\usepackage{graphicx}
\usepackage{dcolumn}
\usepackage{bm}
\usepackage{color}
\begin{document}

\title{Trapping molecules on a chip in traveling potential wells}

\author{Samuel A. Meek}
\affiliation{Fritz-Haber-Institut der Max-Planck-Gesellschaft,
Faradayweg 4-6, 14195 Berlin, Germany}
\author{Hendrick L. Bethlem}
\affiliation{Fritz-Haber-Institut der Max-Planck-Gesellschaft,
Faradayweg 4-6, 14195 Berlin, Germany} \affiliation{Laser Centre
Vrije Universiteit, De Boelelaan 1081, NL-1081HV Amsterdam, The
Netherlands}
\author{Horst Conrad}
\affiliation{Fritz-Haber-Institut der Max-Planck-Gesellschaft,
Faradayweg 4-6, 14195 Berlin, Germany}
\author{Gerard Meijer}
\affiliation{Fritz-Haber-Institut der Max-Planck-Gesellschaft,
Faradayweg 4-6, 14195 Berlin, Germany}

\date{\today}
\date{\today}

\begin{abstract}
A microstructured array of over $1200$ electrodes on a substrate
has been configured to generate an array of local minima of
electric field strength with a periodicity of $120$~$\mu$m
about $25$~$\mu$m above the substrate. By applying sinusoidally
varying potentials to the electrodes, these minima can be made
to move smoothly along the array. Polar molecules in low field
seeking quantum states can be trapped in these traveling
potential wells. This is experimentally demonstrated by
transporting metastable CO molecules in $30$~mK deep wells
that move at constant velocities above the substrate.
\end{abstract}

\pacs{37.10.Mn, 37.10.Pq, 37.20.+j, 33.57.+c}

\maketitle

The manipulation of ultracold atoms above a microchip using
magnetic fields is currently a very active area of research.
Although initially motivated by the strong confinement that
results from miniaturization of guides
and traps, chip technology also enables multiple tools and
devices to be integrated into a single compact structure. Today,
atom chips are therefore discussed in a far more general
context and are, for instance, used for matter-wave
interferometry and precision force sensing~\cite{Fortagh-RMP-2007}.
Via miniaturization, strong electric field gradients can be
produced as well. This has been used to push atoms along a
magnetic guide on a chip~\cite{Kruger-PRL-2003} and to demonstrate
ac electric trapping of atoms in a microscopic volume~\cite{Kishimoto-PRL-2006}.
The strong electric field gradients are also ideal for the precise
manipulation of polar molecules on a chip. Such precision is
required, for instance, for proposed schemes of quantum computation that
use polar molecules as qubits~\cite{DeMille-PRL-2002,Andre-NP-2006}.

Compared to atoms, molecules are more difficult to load and
detect on a chip. The methods to produce the required dense samples
of cold molecules are less matured. As molecules in general lack a
closed two-level system, efficient detection using absorption or
laser induced fluorescence is not possible. These difficulties have
hindered experiments with molecules on chips, and thus far, only reflection
of a slow beam of ammonia from a microstructured electrostatic mirror
has been demonstrated~\cite{Schulz-PRL-2004}.

In this Letter, we present the loading of polar molecules from a
supersonic beam into traveling potential wells on a chip. Metastable
CO molecules are captured in approximately $30$~mK deep wells and
transported at velocities around $300$~m/s along the full $5.0$~cm
length of a chip, and they are detected after leaving it. The operation
principle of this ``supersonic electric conveyor belt'' on a chip is
detailed, and its potential as decelerator is considered.

\begin{figure}[pt]
\includegraphics[height=4.5in]{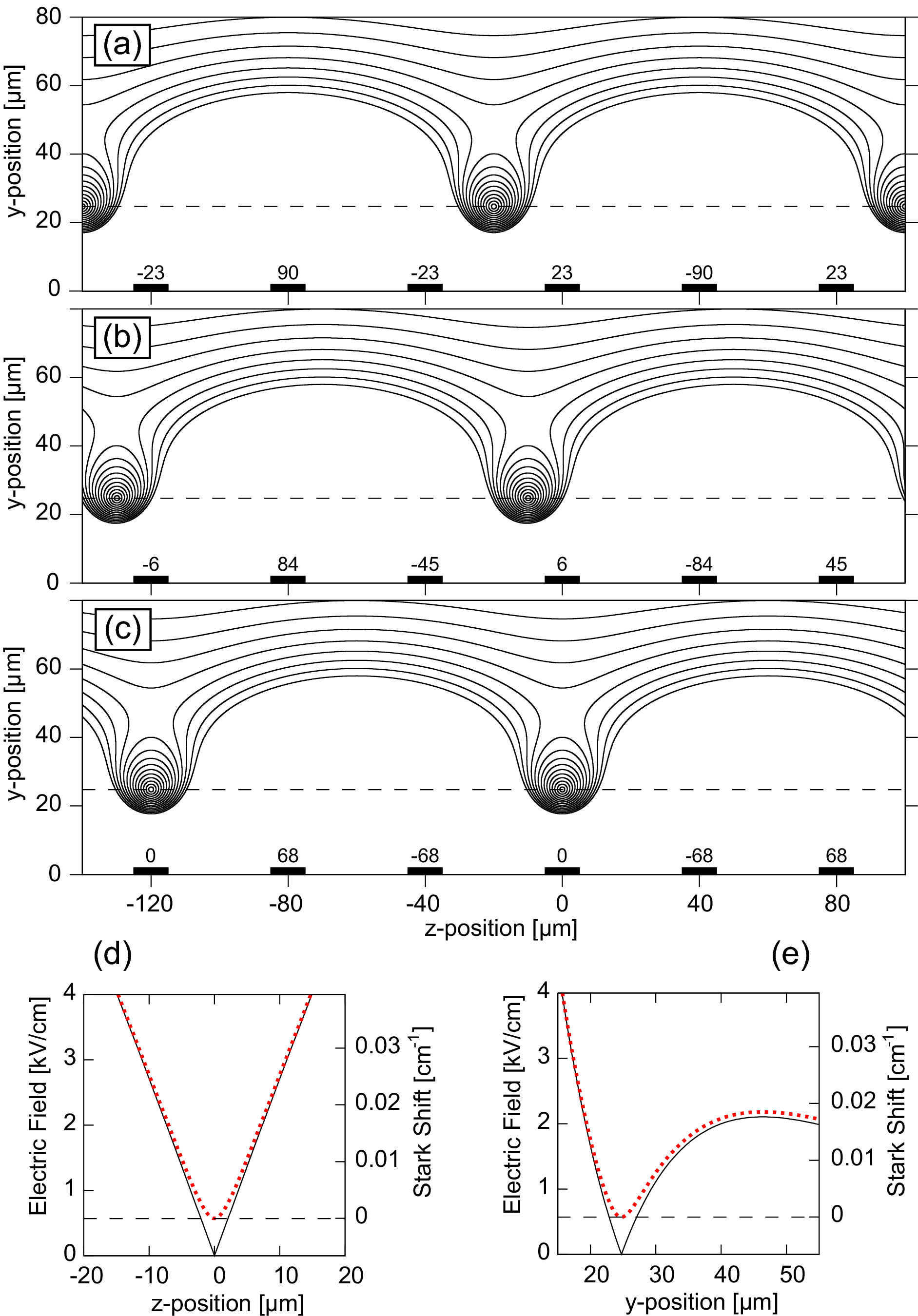}
\caption{((a)-(c)) Contour lines of equal electric field strength
in a plane perpendicular to both the substrate and the electrodes,
shown in $0.2$~kV/cm intervals, for three different times in the
rf cycles. The instantaneous potentials applied to the $10$~$\mu$m
wide electrodes are indicated (in volts). The dashed horizontal lines
are at the same height above the substrate in (a)-(c). The electric field
strength (solid black curve) and the Stark shift for metastable CO molecules
(dashed red curve) are shown around the minima as a function of position
along the $z$ axis (d) and as a function of height above the
substrate (e). Note that the right vertical axes have an offset.} \label{fig1}
\end{figure}

To explain the operation principle of the electric conveyor belt,
we start by discussing the electric field distribution
that can be generated above an array of equidistant, parallel
electrodes. The width of the electrodes is taken to be $10$~$\mu$m
and the center-to-center distance between adjacent electrodes $d$
is $40$~$\mu$m, the parameters used in the experiments described
later. In Figs.~\ref{fig1}(a)-\ref{fig1}(c) contour lines of equal electric
field strength are shown in the $(y,z)$ plane, in $0.2$~kV/cm
intervals. The $y$ axis is perpendicular to the substrate,
whereas the $z$ axis is along the substrate, perpendicular to the
electrodes, which are indicated by horizontal black bars. The
potentials applied to the electrodes are indicated directly
above them (in volts). Calculations are performed for the
two-dimensional case; i.e., the extension of the electrodes
along the $x$ axis is taken to be infinite. Since the actual length of
the electrodes used in the experiments is $4$~mm ($\gg d$), this
approximation is justified. The strength of the
electric field generated above the electrode array generally falls
off rapidly with increasing height above the substrate. At specific
locations above the substrate, however, local minima of the electric
field are created. This is similar to the minima in magnetic field
that can be created above an arrangement of four current carrying
wires~\cite{Dekker-PRL-2000},
and can be most easily understood when inspecting the
electrode arrangement near the center of Fig.~\ref{fig1}(a). With the
potentials applied as shown, the electrodes at $z=0$ and
$z=-40$~$\mu$m create an electric field that is parallel to the
$z$ axis, in the direction of negative $z$ values, for any point
along the vertical line at $z=-20$~$\mu$m. The electrodes at
$z=40$ and $z=-80$~$\mu$m create an electric field that is
pointing in the opposite direction for any point along this line.
For short distances above the substrate, the field due to the two
nearest electrodes dominates, whereas for distances further
away, the field due to the next nearest electrodes is most important.
It is clear, therefore, that at some point above the substrate,
at a typical height on the order of $d$, a zero of the electric
field strength will be generated on the $z=-20$~$\mu$m axis. In this
particular case, this minimum occurs at a height of $25$~$\mu$m
above the substrate.

Rather than using static voltages, the potentials applied to all the
electrodes are sinusoidally varying in time with an angular frequency
$\omega$, which is in the MHz range for our experiments. In each of the panels
(a)-(c) the situation is shown at a given time ($t_a$)-($t_c$),
respectively, with $\omega$($t_b$-$t_a$) = $\omega$($t_c$-$t_b$) = $\pi/6$.
The different panels therefore show the instantaneous voltages
and the resulting electric field distributions at phases of the
rf cycle that are separated by $30^\circ$. The voltage applied to
the $n^{th}$ electrode from the left in Fig.~\ref{fig1}(a) is given by
$-V_0 (1 + cos (\omega t_a + \phi_n))$, with
$V_0 = 45$~V and $(\omega t_a + \phi_n)$ equal to $120^\circ$, $240^\circ$,
and $0^\circ$, for $n=1$, $n=3$, and $n=5$, respectively. For the $2$nd, $4$th, and $6$th electrodes in
Fig.~\ref{fig1}(a), these voltages are $V_0 \cdot (1 + cos (\omega t_a + \phi_n))$,
with $(\omega t_a + \phi_n)$ equal to $0^\circ$, $120^\circ$, and $240^\circ$,
respectively. In this way, the potential on any given electrode
is always equal in magnitude but opposite in sign to the potential
on the electrodes that are three positions
further. The six electrodes shown in Fig.~\ref{fig1}(a) then form one
full period and create two minima of the electric field strength above the
substrate that are separated by $3d$, or $120~\mu$m. Figures \ref{fig1}(b)
and \ref{fig1}(c) are contour plots taken $30^\circ$ and $60^\circ$ later
in the rf cycles, respectively. It can be seen from these three
contour plots that the minima have almost the exact same shape and
are located at an almost constant height above the substrate in any phase
of the rf cycle. The calculated maximum difference in height when the
minima move from the position between two electrodes to a position
centered above an electrode is less than $0.1$~$\mu$m. As a function of time
the minima therefore move smoothly along the substrate, with a velocity
given by $3d(\omega/2\pi)$.

Figures \ref{fig1}(d) and \ref{fig1}(e) show one-dimensional cuts through the electric field
minimum above the electrode centered at $z=0$~$\mu$m in panel (c).
Along the $z$ axis, the electric field strength increases linearly
with distance from the center at $z=0$, like in an ideal quadrupole electric
trap. Along the $y$ axis, the electric field strength has the same quadrupole
behavior close to the minimum. For distances further away from the substrate,
however, there is a slower increase followed by a saddle point of the electric
field strength, limiting the overall depth of this electric trap to about
$2$~kV/cm. The right vertical axes of Figs.~\ref{fig1}(d) and \ref{fig1}(e) indicate the Stark
shift that metastable CO ($a^3\Pi_1, v=0$) molecules in the low-field seeking component
of the $J=1$ level experience in these electric fields~\cite{Jongma-CPL-1997}.
Because of the $\Lambda$ doubling, the Stark shift is quadratic for small values
of the electric field. It is concluded that, above the array of electrodes,
a traveling potential well with a depth of almost
$30$~mK is created for these CO molecules. The shape of this potential
well in the $(y,z)$ plane is approximately circular with a diameter of
about $15$~$\mu$m.

\begin{figure}[pt]
\includegraphics[height=4.445in]{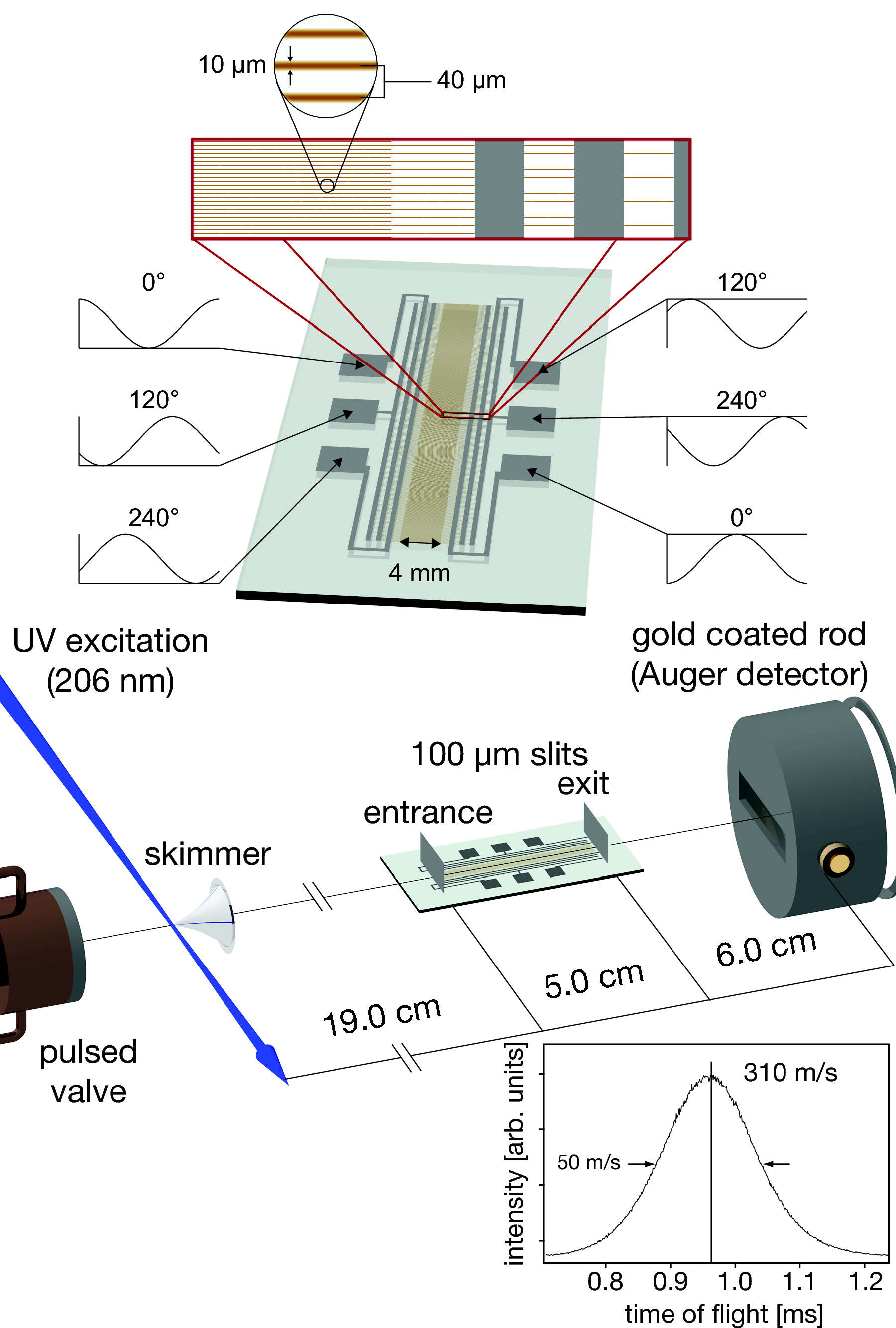}
\caption{Upper part: Scheme of the microstructured electrode array
at different levels of detail. The applied waveforms are schematically
indicated. Lower part: Scheme of the experimental
setup. A pulsed laser prepares the metastable CO molecules in the
desired quantum state. After passing through a skimmer and a
narrow slit the molecules travel closely above the microstructured
electrode array, over its full $5.0$~cm length, and then
pass through a second narrow slit. Time-resolved detection
of the metastable CO molecules takes place another $6$~cm further
downstream. In the lower right corner, the arrival time distribution
of the metastable CO molecules is shown when no voltages are applied
to the electrodes. } \label{fig2}
\end{figure}

The actual implementation of the microstructured electrode array
(micro resist technology GmbH) is shown in the upper part of
Fig.~\ref{fig2}. The $10$~$\mu$m wide and
approximately $0.5$~$\mu$m thick gold electrodes are patterned on a
glass substrate via lithographic etching. In total there are
$1254$ electrodes at a center-to-center distance of $40$~$\mu$m, making for
a total length of the structure of $5.0$~cm. As indicated, the ``active
region'' in which all electrodes are equidistant extends over $4.0$~mm.
The six different time-varying potentials that need to be applied to
the electrode array are coupled in via the three macroscopic pads
that are seen on either side of the structure. The
three stripes that run along the long axis on each side of the
active region and that supply the voltages to every sixth
electrode, are actually at a slightly different height than the
gold electrodes. A $200$~nm
thin protecting and insulating intermediate layer prevents short
circuits of the electrodes when they cross.
This same layer, which has an electrical permittivity of $\epsilon=4$,
also covers the whole microstructured electrode array to reduce the
risk of electrical breakdown. Its influence, as well as that of the
glass substrate, has been taken into account in the calculated
electric field distributions shown in Fig.~\ref{fig1}.

A scheme of the experimental setup is shown in the central part of
Fig.~\ref{fig2}. A pulsed beam ($10$~Hz) is formed by expanding a mixture
of $20$\% CO in xenon into vacuum through a cooled solenoid
valve. In the molecular beam, only the lowest rotational levels in
the electronic and vibrational ground state of CO are occupied, and
molecules in these levels interact only very weakly with external
electric fields. Metastable CO ($a^3\Pi_1, v=0$) molecules, which
experience a strong interaction with electric fields, are prepared
in the low-field seeking component of the $J=1$ level via pulsed laser
excitation on the $Q_2(1)$ transition of the $a^3\Pi \leftarrow X^1\Sigma^+$
band around $206$~nm. The radiative lifetime of this particular
level is $2.63$~ms~\cite{Gilijamse-JCP-2007}, sufficiently long for
the experiments presented here. At the same time, this finite
lifetime permits the recording of some laser induced phosphorescence
signal just after the molecules have passed through the $1.0$~mm diameter
skimmer into the second chamber; i.e., it allows for monitoring
the beam intensity. About $19$ cm downstream from the laser excitation
point, the molecules pass through an approximately $100$~$\mu$m
high entrance slit and then travel closely above the microstructured
electrode array over the full $5.0$~cm length of the array. The
molecular beam axis is thus parallel to the $z$ axis
in Fig.~\ref{fig1}, and it passes over the center of the $4.0$~mm
electrodes. When no voltages are applied to the electrodes, a significant fraction
of the metastable CO molecules that have passed through the entrance
slit also passes through the about $100$~$\mu$m high exit slit. Time-resolved
detection of the metastable CO molecules, which have $6.0$~eV of electronic energy,
is performed by recording the Auger electrons that are released from the
central $4$~mm long area of a gold-coated rod when these molecules
impinge on it. In the lower right corner
of Fig.~\ref{fig2}, the thus recorded arrival time distribution
of the metastable CO molecules over the $30$~cm flight distance is
shown. As the metastable CO molecules are produced by the weakly
focused laser beam ($5$~ns duration) in an about $1$~mm long region
along the $z$ axis, the time on the horizontal axis can be directly
related to the velocity. The peak velocity of the CO molecules is
about $310$~m/s, and with a full-width-at-half-maximum (FWHM) of
$50$~m/s the velocity spread is rather large.

\begin{figure}[pt]
\includegraphics[height=2.112in]{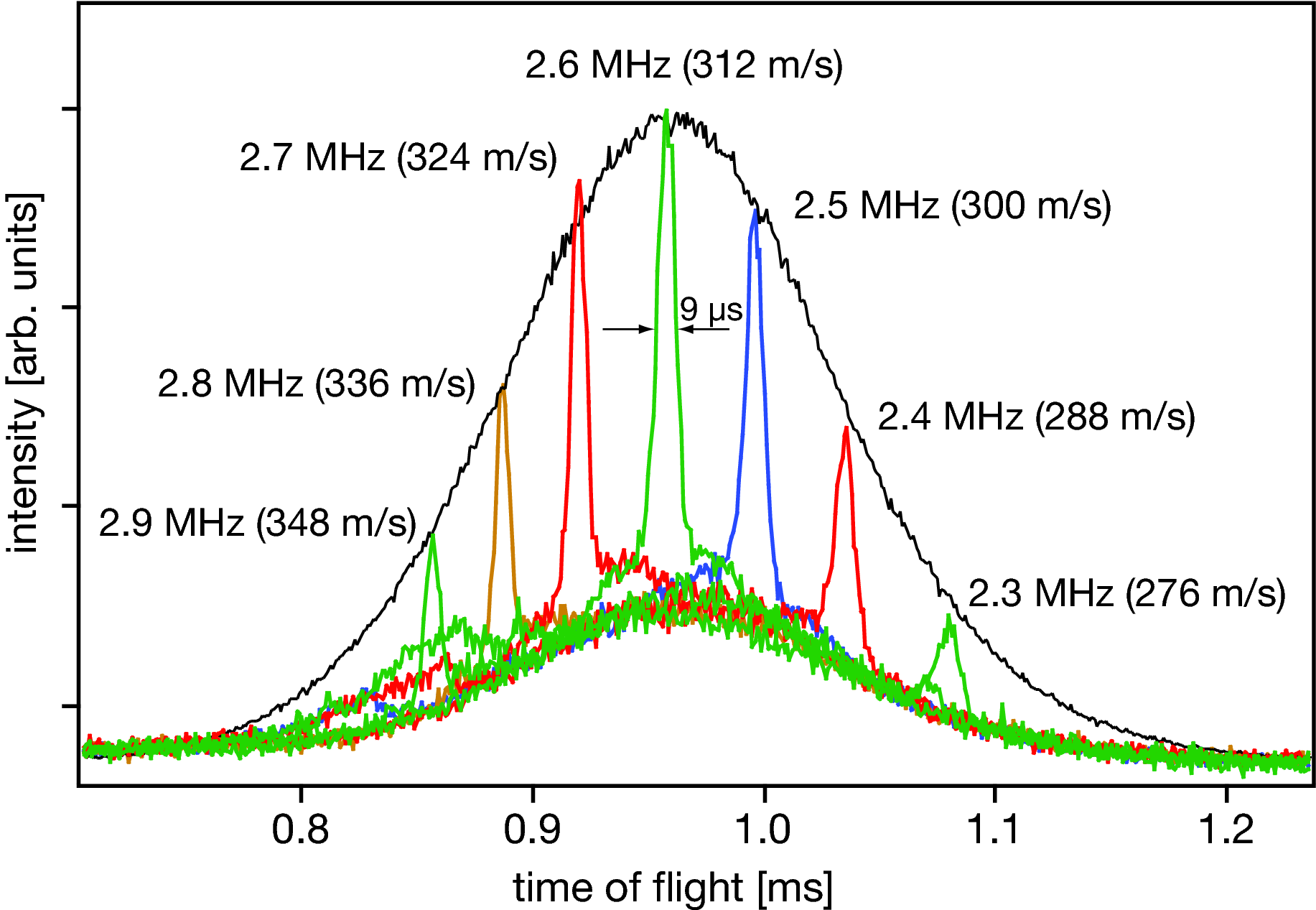}
\caption{Arrival time distributions of metastable CO molecules,
for seven different frequencies $\omega/2\pi$, corresponding
to velocities of the traveling potential wells ranging from
$276$~m/s to $348$~m/s. The enveloping curve is the arrival
time distribution of the molecules in free flight, scaled down
for this comparison. 
} \label{fig3}
\end{figure}

When voltages are applied to the electrodes, the metastable CO
molecules that pass through the entrance slit interact with
the electric fields while flying closely above
the substrate. Most of the molecules are in regions where the
electric field gradient is predominantly perpendicular to the
substrate, and these molecules are efficiently deflected away
from the substrate~\cite{Schulz-PRL-2004}. Consequently, these
molecules no longer pass through the exit slit and will
not be detected. Only the small fraction of the metastable CO
molecules that happens to be near the minima of the electric
field can be transported along the array to the other end. From
purely geometrical arguments, this fraction is expected to be
about $10^{-2}$. In addition, the molecules must have the right
initial velocity to be captured in one of the traveling potential
wells. At a given frequency $\omega$, only an ensemble of molecules
with a FWHM velocity spread of about $5$~m/s around the central
velocity $3d(\omega/2\pi)$ will be accepted.

In Fig.~\ref{fig3}, the measured arrival time distributions of
metastable CO molecules are shown for seven different frequencies
$\omega/2\pi$. The corresponding velocities of the traveling potential
wells are indicated in brackets. For each frequency, a narrow peak
with a FWHM of about $9$~$\mu$s is observed on top of a broad
pedestal. The narrow peaks originate from those metastable molecules
that are actively transported along the microstructured electrode array.
The width of this peak is mainly determined by the velocity spread
of about $5$~m/s, corresponding to a longitudinal translational
temperature of $20$~mK of the sample of guided molecules.
The pedestal might be due to metastable CO molecules
in low-field seeking states that pass high over the substrate
and are not sufficiently deflected. It might also originate
from the small fraction of metastable CO molecules in the
$M = 0$ component of the upper $\Lambda$-doublet level in the
beam, which experiences only a weak interaction with the electric
field~\cite{Jongma-CPL-1997}. The peak intensities of the
transmitted signals for
the different frequencies are determined by the intensities of the
corresponding velocity groups in the original beam. This is
evident from the comparison with the enveloping curve, which shows the
arrival time distribution of the molecules in free flight (scaled
down by more than a factor of $50$ for this comparison).
The integrated intensity of the molecules guided by the $2.6$~MHz 
waveforms is $1$ count per pulse.  With an estimated detector efficiency 
of $1$\%, this represents about $100$ metastable CO molecules from each 
pulse being guided in some $20$ separate minima.

The data presented here demonstrate that polar molecules from a
supersonic molecular beam can be coupled into traveling
potential wells on a chip. This electric conveyor belt is
similar to the magnetic conveyor belt demonstrated for
atoms~\cite{Hansel-PRL-2001}. The six-electrode design
outlined here -- the analogue of which could also be
implemented with magnetic fields -- enables a very
smooth movement of the potential wells over the chip.
By decreasing the frequency of the applied waveforms
while the molecules are on the chip, molecules can be
decelerated and brought to a standstill. Calculations indicate
that metastable CO molecules can be decelerated over a distance
of $40$~mm from $300$~m/s to a standstill, a process that takes
less than $0.3$~ms. The molecules will
then be trapped in stationary potential wells above the
chip. In the present design of the microstructured
electrode array, the molecules are able to escape along
the $x$ axis, but these holes can be plugged using
additional electrodes. Once trapped, the metastable CO
molecules can either be visualized using spatially resolved
imaging of their phosphorescence, or they can be
accelerated off the chip and detected externally.
Alternatively, single molecule detection on a chip using
micro-cavities can be pursued~\cite{Trupke-PRL-2007}.

The operation principle of this Stark decelerator
on a chip is similar to that of the Stark decelerator
demonstrated originally~\cite{Bethlem-PRL-1999}.
However, in that configuration the electric fields are
abruptly switched between two different static configurations,
creating only an \emph{effective} traveling potential
well~\cite{Bethlem-PRL-2000,Gubbels-PRA-2006}. The guiding and deceleration
scheme presented here uses a genuine traveling potential
well. This has the advantage that the molecules only experience
relatively low electric fields. A scaled-up version of this
structure might therefore be useful for decelerating molecules
that are only low-field seeking at low electric fields, such
as heavy diatomics~\cite{Bethlem-JPhysB-2006} and atoms and
molecules in Rydberg states~\cite{Vanhaecke-JPhysB-2005,Softley-JESRP-2005}.

\begin{acknowledgments}
We acknowledge discussions with S.~A. Schulz during the early
stages of this project. The design of the electronics by
G. Heyne, V. Platschkowski, and T. Vetter and the mechanical
design by H. Haak have been crucial for this work.
H.~L.~Bethlem acknowledges financial support from
the Netherlands Organisation for Scientific Research
(NWO) via a VENI-grant.
\end{acknowledgments}

\end{document}